\documentclass[aps,prb,reprint,superscriptaddress,nobalancelastpage,twocolumn,showpacs]{revtex4-1}
\usepackage{color}
\usepackage{comment}
\usepackage{graphicx}
\usepackage{amsmath}
\usepackage{latexsym}
\usepackage{epstopdf}
\usepackage{amsmath,amssymb}
\usepackage{amssymb,amsmath}
\usepackage{multirow}
\usepackage{mathtools}
\usepackage[dvipsnames]{xcolor}
\newcommand{\beq}{\begin{equation}}
\newcommand{\eeq}{\end{equation}}

\usepackage{tikz}
\usepackage{xcolor}
\usetikzlibrary{patterns}
\usepackage{amsthm,amssymb}

\begin{document}
\title{A theorem about time evolution in the quantum mechanics }
\author{ M.~A.~Rajabpour}
\affiliation{  Instituto de F\'isica, Universidade Federal Fluminense, Av. Gal. Milton Tavares de Souza s/n, Gragoat\'a, 24210-346, Niter\'oi, RJ, Brazil}
%\date{\today{}}

\begin{abstract}
Under broad conditions, we prove that the probability amplitudes in the quantum mechanics are either always constant in time
or changing continuously in any interval of time.
\end{abstract}
%\pacs{}
\maketitle

%%%%%%%%%%%%%%%%%%%
%\section{Introduction}

%%%%%%%%%%%%%%%%%%%
In quantum mechanics, the evolution of a state can be described by a unitary operator which can be derived by using Schr\"odinger equation. For time-independent 
Hamiltonians this unitary operator is an exponential function with respect to the Hamiltonian, i.e. $e^{-iHt}$, see \cite{Cohen}. In this note,
we prove a theorem about the evolution of the probability amplitudes in generic many body quantum systems that have lower-bounded Hamiltonians. We also comment about
similar (but restricted) version of the theorem for the evolution of the correlation functions.

The time dependent probability amplitudes for arbitrary states $|\psi\rangle$ and $|\phi\rangle$ is defined as
\begin{equation}\label{Fidelity definition}
F(t):=\langle\psi|e^{iHt}|\phi\rangle.
\end{equation}
When the two states are equal we have the well-known quantum fidelity.

\textbf{Theorem}: For bounded Hamiltonians, if $F(t)=0$ for an arbitrary interval $I=[-\epsilon,\epsilon]$, then, it is identically zero for all the times. 
\begin{proof} The proof is similar to the proof of the Reeh-Schlieder theorem, see \cite{Witten} and references therein. We first define the following quantity for a complex $z$: 
\begin{equation}\label{Fidelity complex}
F(z)=\langle\psi|e^{iHz}|\phi\rangle.
\end{equation}
For Hamiltonians bounded from below, the $F(z)$ is holomorphic in the upper-half plane. Now consider that when we approach to the real $z$,
the $F(z)$ is continuous on the real line and zero in the interval $I$. We will show that this assumption leads to $F(z)=0$.  For $z$ in the upper-half plane, by the Cauchy integral formula, we have
\begin{equation}\label{Cauchy integral formula}
F(z)=\frac{1}{2\pi i}\oint_{\gamma}dz'\frac{F(z')}{z'-z},
\end{equation}
where $\gamma$ is any counterclockwise contour in the upper-half plane. A priory, we do not know that $F(z)$ is holomorphic along the interval $I$, otherwise, by
identity theorem of complex analysis, we could conclude the proof of the theorem. However, since $F(z)$ is vanishing on the interval $I$, one can deform $\gamma$ in a way that part 
of it is along this segment. Then, one can remove this part of the contour and freely move $z$ through the interval $I$ into the lower half-plane.
That means that $F(z)$ is holomorphic also on the interval $I$, it follows that $F(z)$ is identically zero. 
\end{proof}
One consequence of the above theorem is that the fidelity either does not evolve by time or it is continuously evolving in any arbitrary small interval of time.
The following corollary is also an immediate consequence of the above theorem. Consider $F_j(t)=\langle\psi|e^{iHt}|\phi_j\rangle$ then;

\textbf{Corollary I}: For $j\neq k$ and a constant $\alpha$, if $F_j(t)=\alpha F_k(t)$ for an arbitrary interval $I=[-\epsilon,\epsilon]$ then, $F_j(t)=\alpha F_k(t)$ for all the times.

\begin{proof} 
After defining $F_{jk}(z)=F_j(z)-\alpha F_k(z)$ the proof is identical as above.
\end{proof}
The above discussions can not be extended to the equal time correlation functions straightforwardly without further assumptions.  Consider the following
correlation function 
\begin{equation}\label{Correlation function}
\langle A(t)\rangle=\langle\psi| e^{iHt} A e^{-iHt}|\psi\rangle,
\end{equation}
where $A$ can be a product of different operators at different points. The conclusion of the above theorem was based on the holomorphicity of
$\langle\psi| e^{iHt}$ in the upper-half plane, but this is spoiled here by the extra factor $e^{-iHt}$. To be able to use again the holomorphicity argument, we need to assume
that the Hamiltonian is not only bounded from below but also from the above in a way that $|e^{-iHz}|<c$, where $c$ is a finite number. 
With this assumption, one can conclude the following:

\textbf{Corollary II}: For bounded Hamiltonians if $\langle A(t)\rangle=\text{const}$ for an arbitrary interval 
$I=[-\epsilon,\epsilon]$ then, $\langle A(t)\rangle=\text{const}$ for all the times.
\begin{proof} The proof is similar to the proof of the theorem.
\end{proof}
Finally, it is worth mentioning that the above conclusion can be extended  straightforwardly to the mixed states.

\textbf{Acknowledgement}
 I thank J. Dubail and J-M St\'ephan. The work of MAR
was supported in part by CNPq. \\

\end{document}